\documentclass[prl,twocolumn,amsmath,amssymb,floatfix]{revtex4}

\usepackage{graphicx}
\usepackage{bm}
\usepackage{amssymb}
\usepackage{color}
\usepackage{epsfig}
\usepackage{subfigure}

\newcommand{\be}{\begin{equation}}
\newcommand{\ee}{\end{equation}}
\begin{document}

\title {Classical Diffusion of a quantum particle in a noisy environment}

\author{Ariel Amir, Yoav Lahini and Hagai B. Perets}

\affiliation {Faculty of Physics, Weizmann Institute of Science,
Rehovot, 76100, Israel}

\begin{abstract}

We study the spreading of a quantum-mechanical wavepacket in a
one-dimensional tight-binding model with a noisy potential, and
analyze the emergence of classical diffusion from the quantum
dynamics due to decoherence. We consider a finite correlation time
of the noisy environment, and treat the system by utilizing the
separation of fast (dephasing) and slow (diffusion) processes.  We
show that  classical diffusive behavior emerges at long times, and
we calculate analytically the dependence of the classical diffusion
coefficient on the noise magnitude and correlation time. This method
provides a general solution to this problem for arbitrary conditions
of the noisy environment. The results are relevant to a large
variety of physical systems, from electronic transport in solid
state physics, to light transmission in optical devices, diffusion
of excitons, and quantum computation.
\end{abstract}

\maketitle

\emph {Introduction}.- The dynamics of quantum particles, described
by the Schrodinger equation, manifests rich and exciting physics.
The dynamics of a quantum particle in homogenous or periodic media
is ballistic - i.e., the mean-square displacement of a quantum
particle grows linearly with time, in contrast to the diffusive
propagation of a classical particle. In many realistic cases, such
quantum particles are exposed to time-dependent noisy environments,
and decoherence effects become important. In this work we focus on
the emergence of seemingly classical diffusion from the quantum
dynamics under such conditions. Specifically, we calculate the
classical diffusion coefficient as a function of the parameters of
the noise - the magnitude and the correlation time. Our analytical
calculations are found to be in excellent agreement with numerical
simulations.

To formulate the transition from quantum to classical diffusion, we
use a tight-binding model with a stochastic potential. The problem
of the spreading of a quantum wave-packet in a dynamic,
time-dependent potential has received much attention in the last
three decades
\cite{{diffusion_jetp},{post},{kumar1},{kumar2},{Heinrichs},{superdiffusion},{diffusion_lattice},{noise_bound}}.
More recently this problem was discussed in the context of quantum
random walks with decoherence and quantum computation
\cite{kendon,yin+08}. This simplified theoretical problem is of
great relevance to various experimental systems, such as particle
diffusion in molecular crystals \cite{{post}}, diffusion of excitons
\cite{exciton}, photon propagation in coupled waveguide lattices
\cite{{lahini},{Silberberg}}, and destruction of Anderson
localization by nonlinearity or dimensional crossover effects
\cite{pikovsky, kopidakis, flach, Genack}. The problem of dynamics
and dephasing due to coupling to a thermal bath is also closely
related \cite{{holstein},{esposito1},{dubi}}. Early works
\cite{diffusion_jetp} showed that unlike the static disorder case,
where Anderson localization sets in \cite{Anderson}, diffusive
behavior takes over at long enough times for a tight-binding model
with time-dependent disorder. Madhukar and Post [\onlinecite{post}]
extended this to the case of on-site as well as off-diagonal
disorder. Various works
\cite{{kumar1},{superdiffusion},{dissipative_diffusion}} dealt with
the problem on a continuous lattice, which can show superdiffusive
behavior. Heinrichs [\onlinecite{Heinrichs}] showed the
correspondence between discrete (tight-binding) and continuous
scenarios. All of these works assumed, for theoretical simplicity,
delta-function correlations in time (white noise). The corresponding
experimental assumption for electron diffusion in molecular
crystals, for example, is a temperature higher than the Debye
temperature \cite{post}. Clearly, one would like to extend the
theoretical understanding to more realistic cases of finite
correlation times of the noisy environment. A first step in that
direction was undertaken by [\onlinecite {Kitahara}], using a
perturbative analysis in the correlation time $\tau$. In this work
we extend this to arbitrarily large correlation times.  We use a
novel method, relying on the separation of fast and slow processes
(dephasing versus diffusion), to derive accurate analytical
solutions for this problem.

\emph {Model and derivation}.- We consider the Schrodinger equation
for a one-dimensional tight-binding model, taking $\hbar =1$. The
equations governing the process are: \be i \frac{dA_j} {dt} =
T(A_{j+1} + A_{j-1}) + \xi_(j,t)A_j \label {dynamics} ,\ee where
$A_j$ is the amplitude at site $j$ and $T$ is the nearest-neighbor
tunneling. The noise term $\xi(j,t)$ is assumed to be Gaussian and
uncorrelated between different sites, and will be characterized by a
time correlation function~$C$: \be \langle
\xi(i,t')\xi(j,t'+t)\rangle= \delta_{ij}C(t). \label
{correlation}\ee It is clear from the definition that $C(-t)=C(t)$.
The typical time of the decay of $C(t)$ is defined as the
correlation time $\tau$, and the typical magnitude is defined as
$W$, i.e., $C(0)=W^2$. We shall assume that $T \ll W$, from which
follows that the dynamics of the phases of $A_{j+1}$ and $A_{j-1}$
will be driven mainly by the second term of Eq. (\ref{dynamics}). In
the lowest order approximation $A^0$ can be written as: \be A^0_{j}
= |A^0_{j}|e^{-i \phi_j(t)} , \label {zeroth} \ee with $\phi_j(t)=
\phi_j(0) + \int_0^{t} \xi (j,t')dt'$.

Due to the random noise driving it, the phase $e^{-i \phi_j(t)}$
will also be a random variable, with a correlation function
characterized by a correlation time (dephasing time) $\tau_\phi \neq
\tau$. We shall later show that this correlation function will be
related to the diffusion constant of the particle.

The problem of finding the correlation function, defined as
$C_{\phi}(t-t') \equiv \langle e^{-i \phi_j(t)} e^{i \phi_j(t')}
\rangle$, is equivalent to that encountered in the physics of
nuclear magnetic resonance (NMR), where spins lose phase coherence
due to the randomly fluctuating magnetic field created by the other
spins. Plugging in the solution for $\phi_j(t)$, we obtain:

\be C_{\phi}(\Delta t)=\langle  e^{-i \int_0^{\Delta t} \xi(t)dt}
\rangle ,\ee where we have omitted the site index from $\xi(j,t)$
since this calculation is essentially for an independent, arbitrary
site.

Formally, calculating this average is similar to the determination
of Debye-Waller factors \cite{mahan} and yields:

\be C_{\phi}(\Delta t)= ^{lim}_ {\delta t \rightarrow 0}
e^{-\frac{{\delta t}^2}{2} \langle(\sum_k \xi_k )^2 \rangle }= e^{-
\int_0^{\Delta t} (\Delta t - t) C(t) dt}. \label
{exact_dephasing}\ee

This is an exact result for the correlation function of the
dephasing process. A similar result was obtained in
[\onlinecite{dicke_narrowing}], in the context of Dicke narrowing.
When $\Delta t \gg \tau$, we can approximate the integral in the
exponent of Eq. (\ref{exact_dephasing}) by $\Delta t \int_0^{\infty}
C(t) = \beta \Delta t W^2 \tau $, where $\beta$ is non-universal and
depends on the form of $C(t)$. For example, an exponential decay
gives $\beta=1$, while a correlation function decaying linearly to
zero gives $\beta= 1/2$. Therefore in this regime:

\be C_{\phi}(\Delta t) \approx e^{-\beta W^2 \tau \Delta t}. \label
{motional} \ee

Thus for times large compared to $\tau$, the correlation time of the
noise, the correlations of $e^{-i \phi}$ decay exponentially with a
dephasing time $\tau_{\phi} \sim 1/{W^2 \tau}$. This is reminiscent
of the well-known NMR phenomenon of motional narrowing
\cite{slichter}.

For short times, $\Delta t \ll \tau$, we can approximate $C(t)
\approx C(0)=W^2$ in the integral of Eq. (\ref{exact_dephasing}),
and obtain:

\be C_{\phi}(\Delta t) \approx e^{-\frac{W^2}{2} {\Delta t}^2}.
\label {motional2} \ee Remarkably, in this limit the dephasing is
insensitive to the form of the correlation function, and is given by
a Gaussian with $\tau_{\phi}=\sqrt{2}/{W}$.

Let us now proceed to analyze the dynamics of the probability
distribution of the particle, characterized by the set of
probabilities $p_j = |A_j|^2$. It is clear that without the
tunneling term $T$ the probabilities will remain constant, since the
sites are uncoupled. However, due to the noise term, the phase of
the amplitudes become a random variable, which, as we shall now
show, will lead to classical diffusion of the probabilities.

From Eq. (\ref{dynamics}), one obtains the exact relation:

\be \frac{dP_j}{dt}= 2T \rm{Im}
\left[A_j^*A_{j+1}+A_j^*A_{j-1}\right] \label{prob} .\ee

Let us take an ensemble average of Eq. (\ref{prob}). To the zeroth
order approximation, given by Eq. (\ref{zeroth}), the phase of
$A_j$,$A_{j+1}$ and $A_{j-1}$ are independent. This will lead, upon
taking the time-average, to the incorrect result $\langle
\frac{dP_j}{dt} \rangle = 0$. To proceed to the next order $A^1$ in
$T/W$, we have to solve the following first-order, linear
differential equation:

\be i\frac{A^1_j}{dt}-\xi_j A^1_j= T (A^0_{j+1}+A^0_{j-1}) .\ee
Defining an integration factor $\mu_j=e^{i\int_0^t
\xi_j(t')dt'}=e^{i \phi_j(t)}$, this can be rewritten in the form:

\be \frac{d [A^1_j \mu_j]} {dt} = \mu_j \frac{T}{i}
(A^0_{j+1}+A^0_{j-1}) .\ee Upon integration we obtain:

\be A^1_j= A^0_j + \frac{T}{i \mu_j(t)} \int_0^t
\mu_j(t')(A^0_{j+1}+A^0_{j-1}) dt' .\ee

A similar equation holds for $A_{j+1}$ and $A_{j-1}$.

Plugging this into Eq. (\ref{prob}) and taking the ensemble-average
we obtain:

\be \langle \frac{dP_j}{dt} \rangle \approx 2T \rm{Im} \left[{(A^0_j
+A^1_j)}^*(A^0_{j+1}+A^1_{j+1})\right]   + \leftrightarrow , \label
{probs} \ee

where $\leftrightarrow$ denotes identical terms upon the
substitution  $j+1 -> j-1$.

$A^0_j$,$A^0_{j+1}$ and $A^0_{j-1}$ are independent, thus, only
three terms remain in the average. The first of these is:

\be I= \langle 2T^2 \rm{Re}[-{{(A^{0}_j)}^*}\int_0^t
\frac{\mu_{j+1}(t')}{\mu_{j+1}(t)} A^0_{j} dt'] \rangle,\ee

while the other two are of similar form, obtained upon changing
$(j+1) \rightarrow j$ and $j \rightarrow j \pm 1$.

Let us evaluate $I$. Plugging in $A^0_j=P_j(t)e^{-i \phi_j (t)}$,
and the explicit form of $\mu_{j+1}(t)$, we get:

\be I= \langle P_j(t) \int_0^t e^{i [\phi_j(t)-\phi_j(t')]} e^{i
[\phi_{j+1}(t)-\phi_{j+1}(t')]} \rangle. \label
{diffusion_derivation_equation} \ee

Now comes the crux of the matter: the assumption $T \ll W$, already
utilized to allow a perturbative treatment in the leading order of
$T$, also leads to a separation of timescales between the rate of
change of the probabilities and the dephasing. This will be checked
explicitly at the end of the calculation. Under this assumption,
$P_j(t)$ can be taken out of the averaging. Thus $ I= P_j(t) Q ,$
with $Q =\langle \int_0^t e^{i [\phi_j(t)-\phi_j(t')]}
e^{i[\phi_{j+1}(t)-\phi_{j+1}(t')]}\rangle$.

Assuming $t \gg \tau_{\phi}$, we obtain that $Q$ is a constant,
related to the dephasing correlation function $C_{\phi}(\Delta T)$
found before, through the relation $Q= \int_0^\infty C^2_{\phi}(t)$.

Using this result Eq. (\ref{probs}) takes the form:

\be \frac{dP_j}{dt} = 2 T^2 Q [P_{j+1}+P_{j-1}-2P_{j}] ,\ee where we
have omitted the averaging notation. This is the well-known
diffusion master equation.  This implies the (ensemble-averaged)
probabilities will indeed diffuse, with a diffusion coefficient $D=2
T^2 Q$.

Combining this with Eq. (\ref{exact_dephasing}) for the dephasing
correlation function, we obtain our main result:

\be D= 2T^2 \int^{\infty}_0 C^2_{\phi}(t')dt'=2T^2 \int^{\infty}_0
e^{- 2\int_0^{t'} (t' - t) C(t) dt} d t'. \label{main}\ee

For the case of short correlation time $\tau W \ll 1$, we can take $
C_{\phi}(t')$ from Eq. (\ref{motional}), and obtain:

 \be D= \frac{T^2}{\beta W^2 \tau}= \frac{2 T^2}{ \int^{\infty}_{-\infty}C(t)dt}. \label{D1}\ee
This complies with the results of Ovchinnikov
[\onlinecite{diffusion_jetp}], who worked in the limit of $\tau
\longrightarrow 0$ (delta-function correlations of the noise).

The main advantage of our method, other than the clear physical
picture gained for the diffusion process, is that we obtain explicit
analytical results also for the regime of long correlation times, as
well as the crossover between the two regimes. For long correlation
times $\tau W \gg 1$,  Eq. (\ref{motional2}) gives $ C_{\phi}(t')$.
Plugging this into Eq. (\ref{main}) we obtain: \be D= {\frac{\sqrt{2
\pi} T^2}{W}}.\label{D2}\ee The crossover between the two regimes
(short and long noise correlation times), is also described by  Eq.
(\ref{main}). The results depend on the form of the noise
correlation function $C(t)$. Notice that for  $\tau W \sim 1$ Eqs.
(\ref{D1}) and (\ref{D2}) give a result of the order of $T^2 \tau$,
which is where the two limits 'match'.

\begin{figure}[b!]
\includegraphics[width=0.5\textwidth]{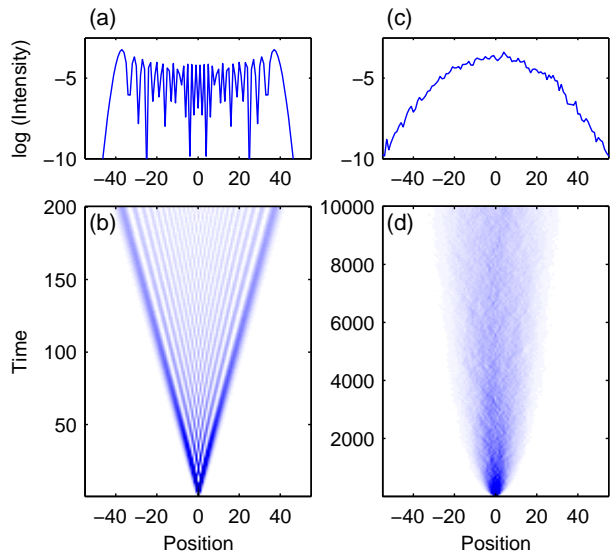}
\caption{The spread of a quantum-mechanical wavepacket in a
perfectly periodic lattice ($T=1$, $W=0$) showing ballistic
spreading (a,b), versus a noisy lattice with $T=1$, $\tau=0.01$,
$W=20$  showing diffusive spread (c,d). The wavefunction is confined
to a single lattice site at t=0. The probability distribution of the
quantum particle after some propagation is plotted on a semilog
scale in (a) and (c), showing a ballistic profile and a diffusive
(gaussian) profile, correspondingly. \label{wavepacket}}
\end{figure}

We have tested the theory numerically, on a finite lattice. The
procedure is as follows:

1.  We solve Eq. (\ref{dynamics}) numerically, starting with a
probability distribution concentrated on a single site, and with a
fluctuating disorder. For convenience, we take noise with linearly
decaying correlations, i.e., $C(t)=W^2
(1-|t|/{\tau})\theta(\tau-|t|)$, where $\theta$ is the Heaviside
step function. The wavepacket spreads, as is illustrated in Fig.
\ref{wavepacket}.

\begin{figure}[b!]
\includegraphics[width=0.4\textwidth]{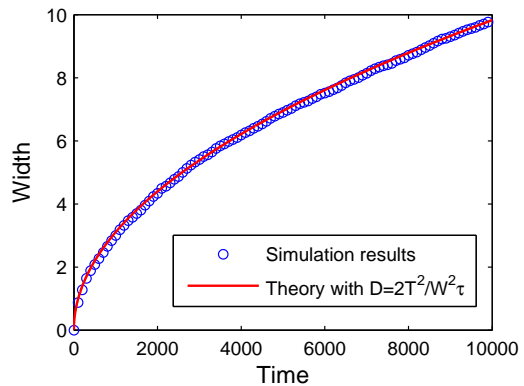}
\caption{ A plot of the width versus time of a quantum mechanical
wavefunction spreading in a noisy potential , showing diffusive
propagation. The simulation parameters are $T=1$, $\tau=0.01$,
$W=20$. The data is averaged over 100 realizations of the disordered
potential. The numerical results (open circles) fit the theory
(line), with no fitting parameters.\label{sqrt_t}}
\end{figure}

2. We ensemble-average the probability distribution over many
realizations of the noisy environment. We find that the standard
deviation of the ensemble-averaged probability distribution grows as
$\sigma(t) \sim \sqrt{2 D t}$, as predicted by the theory. This is
demonstrated in Fig. \ref{sqrt_t}. From this dependence we extract
the diffusion constant $D$. We find indeed that $D$ depends on the
strength of the noise $W$ as well as its correlation time $\tau$.

\begin{figure}[b!]
\includegraphics[width=0.5\textwidth]{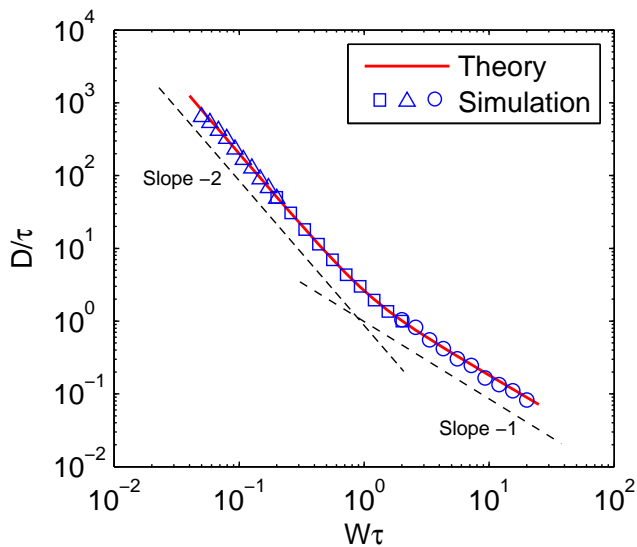}
\caption{A fit of the theoretical results to numerical calculations.
The diffusion constant was found numerically, for correlation times
$\tau=0.01$ (triangles), $\tau=0.1$ (squares) and $\tau=1$ (circles)
and noise magnitudes ranging from $W=2$ to $W=20$.  Each point in
the graph is an average over 50 realizations. As expected
theoretically, there is a crossover from a regime with slope $-2$
(i.e., $1/{x^2}$ dependence), to a regime with slope $-1$ ( $1/{x}$
dependence). The different curves overlap at points with the same
value of $W\tau$. The form of the crossover is given by Eq.
(\ref{main}), integrated numerically for the intermediate regimes.
No fitting parameters are used in the comparison. \label{numerics}}
\end{figure}

Fig. \ref{numerics} compares the result of the simulations with the
theoretical prediction of Eq. (\ref {main}), with a linearly
decaying $C(t)$, as used in the simulations. Since the diffusion
constant is proportional to $T^2$, we know from dimensional analysis
that $D=T^2 \tau f(W \tau)$, where $f(x)$ is determined from Eq.
(\ref{main}). This allows for a unified presentation of both
regimes, which are manifested as different asymptotic regimes of the
function $f(x)$: for $W \tau \ll 1$ we know from Eq. (\ref{D1}) that
$f(x) \sim 1/{x^2}$ with a non-universal proportionality constant,
while for $W \tau \gg 1$ we have $f(x)=\sqrt{2 \pi}/{x}$, as deduced
from Eq. (\ref{D2}). Indeed, we numerically obtain scaling of this
form, where the data for different runs collapse onto a single
curve, when scaled correctly. Fig. \ref{numerics} shows $f(x)$
derived numerically, and compares it to the prediction of Eq.
(\ref{main}). Note that there are no fitting parameters. The
excellent fit confirms the validity of the analytical approach used.

\emph {Summary}.- We have considered a tight-binding model with
random on-site energies, fluctuating in time with an arbitrary
correlation time. Using the separation of timescales between the
(slow) diffusion process and the (fast) dephasing, we managed to map
the complex evolution of the probabilities onto the problem of
dephasing of a \emph{single} site. For a noisy environment with
short correlation time (compared to the reciprocal strength of the
disorder, $1/W$), the phenomenon of motional narrowing is manifested
in the diffusion constant. In the regime of long correlation time,
we show that the diffusion constant is inversely proportional to the
strength of the disorder. We also give a solution for the crossover
regime. Our results show excellent agreement with simulations,
inadvertently confirming previous results obtained for the some
particular cases studied in the past, but continuing much beyond to
give a general solution for the full range of the parameter space.
This includes quantum systems under more realistic noise
environments conditions, which have not been studied before. Our
results and exact solutions are applicable to various quantum
systems, from diffusion of electrons and excitons to photon
propagation in coupled waveguides, and to the general question of
quantum random walks with decoherence.

We dedicate this work to the late Y. Levinson, who contributed to it
by illuminating discussions, and pointed out the striking
resemblance to the physics of motional narrowing. We would like to
thank Y. Oreg and Y. Imry for useful discussions. A.A. acknowledges
funding by the Israel Ministry of Science and Technology via the
Eshkol scholarship program. Y.L. acknowledges support by the Israel
Academy of Science and Humanities via the Adams scholarship program.

\end{document}